# Seeing the gravitational wave universe

## Pulsar timing arrays will be a window into the gravitational wave background

*By* **Chiara M. F. Mingarelli**[1,2] *and* **J. Andrew Casey-Clyde**[1]

[1]Department of Physics, University of Connecticut, Storrs, CT 06269-3046, USA. [2]Center for Computational Astrophysics, Flatiron Institute, New York, NY 10010, USA. Email: chiara.mingarelli@uconn.edu

Gravitational waves are ripples in the fabric of spacetime that are caused by events such as the merging of black holes. In principle, many types of events occur that could create gravitational waves with frequencies ranging from as high as a few kilohertz to as low as a few nanohertz. Sources of gravitational waves in the nanohertz frequency range include cosmic strings, quantum fluctuations from the early Universe, and, notably, supermassive black hole binaries (SMBHBs). Some gravitational wave sources are so numerous that they are all expected to contribute to a gravitational wave background (GWB). This GWB has been the target of pulsar timing arrays (PTAs) for decades.

PTAs use the correlations between dozens of pulsar pairs to observe the GWB. Recently, the North American Nanohertz Observatory for Gravitational Waves (NANOGrav) (*1*), the European Pulsar Timing Array (EPTA) (*2*), the Parkes Pulsar Timing Array (PPTA) (*3*), and the International Pulsar Timing Array (IPTA) (*4*) have all detected a low-frequency noise in their pulsar data, which may be the first hint of the GWB (see the figure).

The common, low-frequency noise (also called red noise) that the PTAs have measured could be due to the cosmic population of slowly evolving SMBHBs. These SMBHBs create gravitational waves with periods of years to decades in their inspiral phase, the time in the binary's evolution leading to the final merger. This inspiral time scale is very long: A typical equal-mass ($1 \times 10^9$ solar mass) SMBHB observed with a frequency of 1 nHz is 25 million years from merging. Indeed, these mergers take so long that they should create a stochastic (or random) GWB as a result of the incoherent superposition of potentially tens of thousands of gravitational wave signals. The GWB signal induces delays and advances in the time that it takes for pulses from millisecond pulsars to reach Earth. This signal can be extracted by cross-correlating the residuals—the difference between the expected and the actual arrival time—of pulsar pairs in the PTA. The noise in each pulsar should be independent, whereas the GWB signal should be a common signal in each pulsar—hence, the more pulsar pairs that can be observed, the lower the noise and the larger the signal. The smoking gun of the GWB is the Hellings and Downs curve (*5*), for which we expect the recently detected red noise to eventually conform to a specific functional form (see the figure).

Although strong evidence exists for a common red-noise process (or low-frequency signal) in all the NANOGrav, PPTA, EPTA, and IPTA pulsars, little evidence has been found so far for the Hellings and Downs curve. Whereas Goncharov *et al.* (*6*) concluded after a series of simulations that some common, or similar, red noise originating in pulsars could mimic the common red noise generated by a GWB, Romano *et al.* (*7*) showed that the detection of a common red-noise process should be expected before the Hellings and Downs spatial correlations. If the correlated red noise that is seen in all PTAs truly is a GWB, then detection should be expected with two to five more years of data (*8*).

Notably, a nanohertz GWB sourced by SMBHBs would indicate that the long-standing final parsec problem—where the SMBHs stall at 1 pc of separation before they efficiently emit gravitational waves—is solved. Having the system stall at a ~1-pc gap would be almost completely ruled out because the gap depletes the GWB amplitude by ~30% (*9*). Indeed, a GWB amplitude commensurate with the current red noise is so large that it would rule out all but the most optimistic GWB models with no such stalling at 1 pc. For example, Casey-Clyde *et al*. (*10*) found that the number density of SMBHBs in a NANOGrav-like GWB would be five times larger than that in the one predicted by Mingarelli *et al*. (*11*). This either signifies that Mingarelli *et al*. (*11*) were too conservative in their mass and merger rate estimates or that perhaps the merger models need an additional level of sophistication, for example, gas and binary eccentricity, which could in turn increase the number of expected SMBHBs.

Although the focus is on SMBHBs because of their expected presence in the PTA frequency range, other sources are possible. A network of cosmic strings, the existence of which has never been directly demonstrated, is another potential source of a GWB. A third source, a GWB of primordial origin, would provide evidence of an ekpyrotic Universe, where the Big Bang is eventually followed by a Big Crunch. It is not known for sure how long it will take to distinguish between different sources, but Pol *et al*. (*8*) showed that at the time of an initial detection of spatial correlations in pulsar pairs with a signal-to-noise ratio of three, current PTAs should have the capability to distinguish a SMBHB from at least some such exotic sources.

Once the GWB is detected, the next task is to make maps of it, akin to the cosmic microwave background. For instance, individual nearby SMBHB systems and potentially large-scale structures could contribute to or trace the anisotropy in the GWB (*12*). Indeed, GWB anisotropy may enable us to constrain the cosmic population of SMBHBs. Moreover, it will be interesting to see where the anisotropic (excess) power on the sky originates and whether this can be associated with SMBHB activity. However, obtaining upper-limit maps of GWB anisotropy may be challenging because the distribution of pulsars in the sky is itself anisotropic, thwarting the use of the usual spherical harmonics (*13*).

Counterintuitively, detecting continuous gravitational waves from individual inspiraling SMBHB systems by PTAs is possible but more challenging than detecting the GWB. All-sky searches for these continuous waves are computationally expensive and provide poor sky localization for detections. A different path forward is to follow up on binary candidates from electromagnetic surveys, which search for periodic light curves, such as the Catalina Real-time Transient Survey (CRTS). Indeed, recent hydrodynamical simulations predict that periodic light curves could roughly trace the binary's orbit [e.g., (*14*)]. Some of these periodic light curves might just be noise that, on short time scales, appears to be periodic. Targeted searches for these binaries appear to be the most promising path forward, because knowing the sky position and rough guess of the binary's period improves PTA sensitivity by an order of magnitude (*15*). As such, extensions to CRTS

## Gravitational wave background

Events such as the merging of supermassive black holes would contribute to a gravitational wave background (GWB) that is potentially detectable by using many pairs of pulsars.

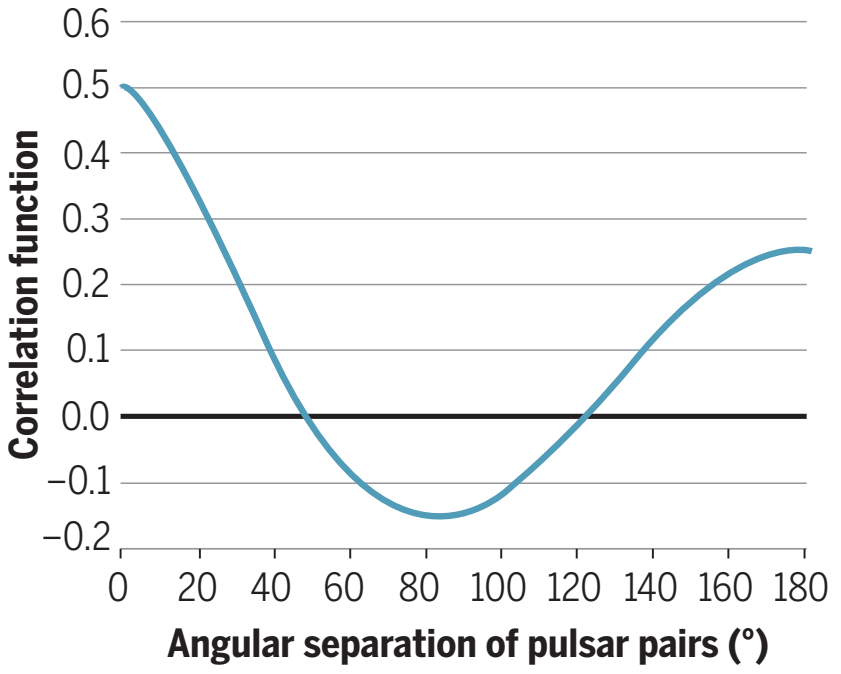

The expected correlation pattern induced by a GWB is called the **Hellings and Downs curve**, which has this specific functional form. Each pulsar pair will appear as a single point on this correlation curve; hence, a credible detection requires a vast number of pulsar pairs. At present, very little evidence of this curve exists in any published pulsar timing array (PTA) dataset.

## Common process

Hints of this background show up as low-frequency noise, found in PTAs. Evidence for a common process (CP) red-noise signal in PTA data is highlighted by data release (DR) 2 from the International PTA (IPTA), which incorporates only 9 years of North American Nanohertz Observatory for Gravitational Waves (NANOGrav) data. Combining European PTA (EPTA) and Parkes PTA (PPTA) data is equivalent to getting three additional years of NANOGrav data.

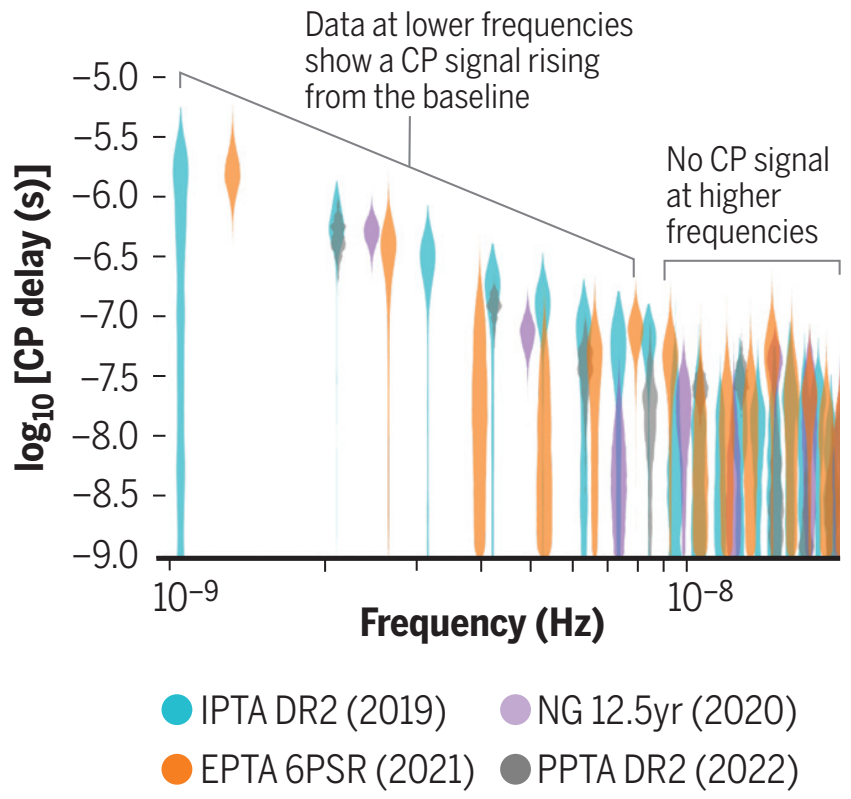

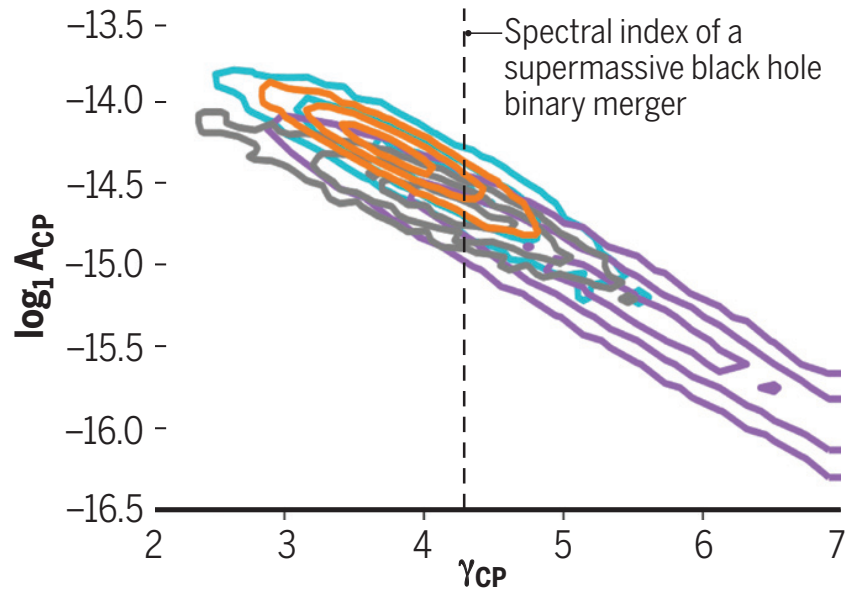

A GWB from supermassive black hole binary mergers should have a spectral index ($\gamma_{cp}$) of 4.33. Identifying both amplitude and spectral index is key to identifying a common process.

and the future Rubin Observatory will be crucial for finding possible electromagnetic counterparts to the SMBHB mergers, and facilities such as the next-generation Very Large Array (ngVLA) will be critical for imaging nearby gravitational wave host galaxies.

The detection of the GWB may be imminent, and as such, a new low-frequency era of GW astronomy is at hand. Assuming that the GWB is
astrophysical, its detection will likely cast aside any remaining doubt that SMBHs do eventually merge. Moreover, it will yield insights into the expected number density of SMBHBs as a function of redshift, the volume enclosing the GWB, and the minimum mass of a SMBHB that contributes to the background (*10*). All these values are fundamental properties of SMBHBs on which there are extremely limited observational constraints (which also come from PTAs). At present, PTA datasets span about 15 years, and with 5 more years of data, it should be possible to measure a low-frequency turnover in the GWB strain spectrum due to the presence of, for example, gas and stars surrounding the cosmic population of SMBHBs (*8*). Underlying all of this exciting astrophysics will be IPTA datasets formed by combining data from all the major PTAs, substantially increasing detection prospects for all nanohertz gravitational wave sources.

**ACKNOWLEDGMENTS**

We thank P. Baker, S. Chen, J. Lazio, D. Nice, M. McLaughlin, N. Pol, J. Romano, S. Taylor, and S. Vigeland for useful comments. This research was supported in part by the National Science Foundation under grant nos. NSF PHY-1748958, PHY-2020265, and AST-2106552. The Flatiron Institute is supported by the Simons Foundation.